# Electrical detection of microwave assisted magnetization reversal by spin pumping

Siddharth Rao, Sankha Subhra Mukherjee, Mehrdad Elyasi, Charanjit Singh Bhatia, and Hyunsoo Yang[a]

*Department of Electrical and Computer Engineering and NUSNNI, National University of Singapore, 117576, Singapore*

Microwave assisted magnetization reversal has been investigated in a bilayer system of Pt/ferromagnet by detecting a change in the polarity of the spin pumping signal. The reversal process is studied in two material systems, Pt/CoFeB and Pt/NiFe, for different aspect ratios. The onset of the switching behavior is indicated by a sharp transition in the spin pumping voltage. At a threshold value of the external field, the switching process changes from partial to full reversal with increasing microwave power. The proposed method provides a simple way to detect microwave assisted magnetization reversal.

a) E-mail: eleyang@nus.edu.sg

The demand for ultrahigh density recording media with areal densities beyond 1 Tb/in$^2$ has led to intensified research in energy assisted recording technologies, such as heat assisted magnetic recording (HAMR)[1,2] and microwave assisted magnetic recording (MAMR)[3,4], since conventional recording techniques are not suitable to generate the necessary magnetic fields to write on high coercivity materials. MAMR has been recently investigated to be feasible on perpendicular magnetic anisotropy media by numerical simulations[4,5] and experiments. Lu et al.[6] reported clear microwave assisted switching at densities upto 700 Gbits/in$^2$, while Boone et al.[7] observed microwave assisted switching in granular media by anomalous Hall effect measurements. In this phenomenon, the microwaves excite the large-angle spin precession creating a reduced magnetic energy barrier for the magnetization reversal. This process is most efficient when the microwave frequency is close to the natural ferromagnetic resonance (FMR) frequency, as the precession angle is the maximum at resonance. Conventionally, FMR and magneto-optical Kerr effect spectroscopy techniques have been the most commonly used to study MAMR. In both techniques, complex impedance matching between the sample and the measurement equipment is required in order to obtain good signals. The FMR voltage can also be detected by electrical methods using the photovoltage (PV) generated by spin rectification effects[8-10], and this method has been used to characterize MAMR in permalloy microstrips[11]. However, this technique is limited by the strength of the anisotropic magnetoresistance (AMR) signal in the investigated materials. Therefore, a characterization technique that can convert the time-varying magnetic signal to a simple output signal for a wide range of materials is of significant importance and timely.

In this letter, we demonstrate a technique of detecting microwave assisted reversal in patterned and thin film magnetic structures utilizing spin pumping phenomena. The spin current pumped into the nonmagnetic layer (Pt) due to magnetization dynamics either in CoFeB or NiFe is converted to a charge current by the inverse spin Hall effect (ISHE), and this charge current is measured as a dc voltage[12-15]. This dc voltage may be primarily attributed to the spin pumping phenomenon in our experiments, as will be discussed. We show that the magnetization reversal of the ferromagnetic material can be detected by a change in the polarity of the measured spin pumping dc voltage signal. Thus, spin pumping offers a simple useful method to study the magnetization reversal mechanism by static measurements. By varying the microwave power at a constant magnetic bias field, we can also analyze the features of the switching process in both patterned and thin film structures.

The films were deposited in a magnetron sputtering chamber with a background pressure of $2\times10^{-9}$ Torr, followed by standard photolithography, ion milling, and lift-off process steps. Two samples with in-plane anisotropy were prepared. Sample A was a Pt (10 nm)/$Co_{60}Fe_{20}B_{20}$ (20 nm) stack, and sample B was a Pt (10 nm)/$Ni_{78}Fe_{22}$ (Py, 20 nm) stack. A Pt strip of dimensions 800 $\mu$m × 600 $\mu$m × 14 nm is patterned first, followed by addition of the magnetic layer, before which 4 nm of the Pt stack is ion-milled to ensure good electrical contact between the two layers. The magnetic layer in Sample A was patterned into a large thin film of dimensions 300 $\mu$m × 619 $\mu$m × 20 nm as shown in Fig. 1(a). Sample B, on the other hand, was patterned into an array of 53 equally-spaced microwires of dimensions 5 $\mu$m × 619 $\mu$m × 20 nm with a spacing of 5 $\mu$m, as shown in Fig. 2(a). Both sample stacks were encapsulated by a 20 nm thick $SiO_2$ layer covering only the area of the magnetic layer. Finally, shorted coplanar waveguides (CPW) and dc probe pads of Ta (5 nm)/Cu (150 nm) were defined and sputter-deposited on the samples. The oxide layer between the CPW and the magnetic layer ensures that only the microwave field is coupled to the latter and there is no current shunting effect. Sample A was annealed at 300 °C for 30 minutes in a magnetic field of 200 Oe.

Figure 1(a) also shows a schematic representation of the experimental setup and the orientation of the co-ordinate axes. The bias field ($H_b$) was applied along the *x*-axis, and the microwave signal was applied to the 50 Ω-shorted CPW to generate an Oersted field **h** perpendicular to the external bias field. This setting (**h** ⊥ $H_b$) induces precessional dynamics in the magnetic layer. The precessing magnetization pumps a spin current into the adjacent Pt layer, and this spin current generates a proportional charge current across the Pt strip as a result of ISHE. The charge current was detected as a dc voltage at the contact pads. The measured dc voltage in this experiment can be due to two effects, such as the spin pumping effect and spin rectification effects[8,10]. Due to capacitive coupling across the oxide layer, a small portion of the microwave current may be induced in the ferromagnetic layer. This current flow in the ferromagnetic layer may generate an additional dc voltage due to anisotropic magnetoresistance (AMR)[16,17], anomalous Hall effect (AHE)[18,19] and planar Hall effects (PHE)[19].

In order to evaluate the contributions of spin pumping and spin rectification effects, the measured voltage was fitted with a Lorentzian wave-function to extract the symmetric and antisymmetric components. Based on previous experiments[16-19], it is known that the AMR effect can induce a dc voltage with both symmetric and antisymmetric components, while the dc voltage due to the AHE effect has an antisymmetric line-shape. We find that the antisymmetric component (0.42 $\mu$V) is one order lesser than the symmetric component (4.4 $\mu$V) and hence, is of negligible effect in our experiments. To analyze the contributions to the symmetric component of the measured voltage, an angular dependence study was performed by varying the angle ($\theta$) of the magnetic bias field with respect to the direction of the microwave current. The measured data showed a clear dependence on the $\cos\theta$ function, which is known to be the characteristic of the spin pumping phenomenon[17,20]. The above analysis suggests that the spin pumping effect is the plausible and dominant mechanism for generating the dc voltage in our samples and measurement configuration. However, a technique to generally separate the spin pumping signal from the rectification effects due to AHE and PHE has not been well-resolved so far. Therefore, our samples may also have a dc voltage due to rectification effects.

The basic magnetic properties of the film were analyzed by studying the spin pumping signal as a function of the bias field, as shown in Fig. 1(b). The sample is initially saturated by a strong bias field in the $-x$ direction. The microwave signal is applied in a frequency range from 1 to 10 GHz at a constant power of 15 dBm, while the bias field is swept across the resonance field, given by the Kittel formula $f_{res} = \gamma\sqrt{H_{res}(H_{res}+M_s)}$, where $\gamma = 2\pi(28$ GHz/T$)$ is the gyromagnetic ratio. The magnetic layer is reset to its initial state for every microwave frequency. At resonance, the number of spins pumped into the Pt layer is maximized, therefore a peak is observed in the output voltage in Fig. 1(b). The peak shape is a summation of the symmetric and anti-symmetric part of the spin pumping signal, and the resonance field ($H_{res}$) can be extracted by fitting the waveform to a Lorentzian wavefunction. The saturation magnetization ($M_s$) of the deposited CoFeB film in sample A is then extracted by fitting the resonance frequency versus bias field data to the Kittel formula, as shown in Fig. 1(c). The $M_s$ for sample A was estimated to be 2.07 T. The FMR line-widths (62 Oe at 5 GHz) are also comparable to the previous reports[21], and the enhancement of Gilbert damping due to spin pumping has been observed similar to others[22-26].

As the microwave field increases, the precessional torque and as a result, the precession angle increases under conditions of resonance. If the rate of energy absorption from the microwaves exceeds the rate of damping, the magnetization will reverse for a sufficient value of the bias field. This reversal is detected by a change in the polarity of the ISHE voltage, as the orientation of the pumped spins has changed. The charge current, generated due to ISHE, is related to the pumped spin current as $J_c \propto J_s \times \sigma$, where $J_c$ is the charge current density, $J_s$ is the density of spin current induced from spin pumping, and $\sigma$ is the spin orientation. A change in the spin orientation will cause a change in the direction of charge current flow. Therefore, we see a change in the polarity of the ISHE voltage upon magnetization reversal. The two main tuning parameters of the microwave signal are the excitation power ($P_{inp}$) and frequency ($f$). Figure 1(d) shows the results of the variation in microwave power. At $P_{inp}$ = 13 dBm, the magnetization shows a clear resonant behavior (a dip) for $H_b$ = 4.1 Oe and 15.6 Oe, before it completely reverses at $H_b$ = 27.2 Oe resulting in a peak. At $P_{inp}$ = 14 dBm, the magnetization exhibits an abrupt change rather than a broad dip in the measured spin-pumping voltage ($V_{SP}$) at $H_b$ = 15.6 Oe. The sudden decrease and change in the polarity of $V_{SP}$ can be attributed to the partial reversal of the magnetic layer (domains are formed in parts of the CoFeB thin film), while the rest of the film remains in its initial magnetic state. This reversal causes a change in the orientation of some of the pumped spins, and consequently the overall ISHE voltage decreases in its absolute value. As the microwave power increases to 16 dBm and then 18 dBm, this domain reversal behavior extends to larger regions. The stronger microwave fields cause larger domains to form, and at 18 dBm the reversal is almost complete as evidenced by the presence of a clear peak with reversed polarity and line-widths comparable to those at higher bias fields. The increasing value of $V_{SP}$ at higher microwave powers is due to larger oscillations causing higher spin pumping and in turn, a higher spin current that generates a higher charge current. At $P_{inp}$ = 20 dBm, the magnetization has completely reversed from a dip to peak at $H_b$ = 15.6 Oe, as there are no visible abrupt changes in the output signal.

In order to demonstrate that this characterization technique is suitable for other materials, we have studied a patterned array of Py microwires in sample B. The external bias field was applied along the axis of the wires ($x$-direction) and the microwave field is applied perpendicular to the axis in Fig. 2(a). The film properties of sample B were characterized by measuring $V_{SP}$ as a function of the bias field as shown in Fig. 2(b) and the extracted $M_s$ from a fit, shown in Fig. 2(c), is 0.81 T. The reversal behavior was studied by sweeping the microwave power from 0 to 20 dBm, in finer steps of 1 dBm due to the soft magnetic nature of Py. Figure 2(d) shows a characteristic data set at 4 dBm intervals where the switching is most evident. At $P_{inp}$ = 4 dBm, magnetization reversal is seen at $H_b$ = 20.3 Oe indicated by a change from a dip to peak. As the power increases to 8 dBm, we see a small change at $H_b$ = 11.8 Oe and $f$ = 1.75 GHz that is likely a case of partial switching followed by relaxation to initial state. At 12 dBm, this reversal is more prominent at $H_b$ = 11.8 Oe represented by a broad peak, however at $f$ = 1.75 GHz the magnetization suddenly reverses back to its initial state. At higher powers, the Py wires switch completely and remain stable in the new configuration (peak).

The random 'switch-back' events at $P_{inp}$ = 12 dBm can be attributed to the effects of spin pumping into the Pt strip and the design of the CPW over the microwire array. The dimensions of the CPW in both samples are 270 $\mu$m × 619 $\mu$m which covers 90 % of the magnetic layer in sample A and 50% of the magnetic layer in sample B. As a result, the microwave field strength decreases as the distance of the wire from the CPW increases[27]. As the microwave power increases, the wires underneath the CPW undergo reversal, while those away from the CPW do not switch as the microwave-generated fields are not strong enough to influence magnetization reversal. This partial switching is a likely scenario as seen from the output voltage frequency response at $P_{inp}$ = 12 dBm where two features at 1.25 GHz (peak) and 1.75 GHz (dip) are seen. At higher microwave powers when the microwave field *h* is strong enough to overcome these inhomogeneous field effects, complete switching happens.

In order to gain a more qualitative picture of the MAMR process, we look at the switching process as a function of the applied microwave power at a constant value for $H_b$. Figures 3(a-c) show the data for an applied power in the range of 11 – 20 dBm in the case of CoFeB sample. The CoFeB film was saturated in the –*x* direction for every increment of the microwave power. At $H_b$ = 3.3 Oe in Fig. 3(a), no switching occurs as the microwave field at $P_{inp}$ = 20 dBm is not high enough to switch the material. At $H_b$ = 15.2 Oe in Fig. 3(b), we observe the onset of microwave-assisted reversal at $P_{inp}$ = 14 dBm, and by 19 dBm the material is completely switched indicated by a change from a dip to peak. At $H_b$ = 39 Oe in Fig. 3(c), the reversed polarity of $V_{SP}$ indicates that the magnetic layer has already switched at $P_{inp}$ = 11 dBm. A similar trend in the switching profile is observed for sample B (NiFe) at the static field of $H_b$ = 10.5 Oe. In this case as shown in Fig. 3(e), partial microwave-assisted switching is seen at $P_{inp}$ = 11.57 dBm. No switching occurs for $H_b$ = 8.1 Oe in Fig. 3(d), while the static field contributions at $H_b$ = 16.4 Oe in Fig. 3(f) is strong enough to switch below $P_{inp}$ = 5 dBm. In comparison to characterization methods based purely on spin rectification effects such as the photovoltage (PV) effect[11], the spin pumping-based characterization technique is not limited by the strength of the AMR effect in the ferromagnetic layer. While the PV technique is a simpler dc measurement technique as compared to the spin pumping technique, the qualitative information about partial switching in the latter makes it an attractive proposition for testing MAMR in arrays of magnetic patterns, as in a storage media disk.

The reversal mechanism can be discussed as a result of the competition between the injected energy and damping energy in a system with a microwave excitation. The external static bias field reduces the energy barrier height between the two stable magnetization states, and the absorption of energy from the microwave field causes the precessing magnetization to climb up the energy well and arrive at the threshold point. Considering a perturbative approach[28], the magnetization in a single domain is assumed to be precessing with no damping as it is taken into account as a separate energy contribution. If the magnetization is precessing with the same angular frequency as the microwave excitation in a given precessional mode, then the reversal is determined by a competition between the injected and damping energies. If the injection process is dominant, a particular amount of energy is coupled to spin wave excitations[29,30] and this makes the precessional mode unstable resulting in a change of the precessional path to a different cone angle. Depending on the values of the Gibbs free energy, the injected, and damping energies of the system, the magnetization will surmount the energy barrier and fall in the other energy well, thereby achieving magnetization reversal. In a multi-domain model, the effect of microwaves on magnetization reversal is initiated by domain nucleation as seen in our experiments. This is followed by a reversal process similar to the single domain model, but with additional interactions between domains due to spin wave modulations and domain wall dynamics, which requires more in-depth research to quantify their contributions.

In summary, we have presented a technique of characterizing MAMR by utilizing the spin pumping phenomenon. This technique is able to provide a quick and accurate picture of the reversal mechanism in the magnetic structure, regardless of material parameters and geometric features. A qualitative understanding of the factors governing the reversal mechanism, specifically the competition between the injected and damping energies was presented for both single domain and multi-domain models.

This work was supported by the Singapore NRF CRP Award No. NRF-CRP 4-2008-06.

**FIGURE CAPTIONS**

FIG. 1. (a) Schematic representation of the device geometry (not to scale) and the measurement setup. The CPW is connected to a signal generator (SG) and a voltmeter is connected across the Pt for measuring the spin-pumping signal ($V_{SP}$). The lower part shows the cross-sectional view and film composition of sample A. The CPW shown here is a simplified version. (b) $V_{SP}$ as a function of the bias field ($H_b$) at a constant microwave of 15 dBm, with an offset for clarity. (c) Kittel fitting of the resonance peaks to extract the saturation magnetization ($M_s$). (d) Variation in $V_{SP}$ as a function of magnetic field for different microwave excitation powers, with an offset for clarity.

FIG. 2. (a) Cross section view and film composition of sample B. The magnetic layer is permalloy (NiFe) patterned into an array of 5 $\mu$m-wide wires with a spacing of 5 µm. (b) Spin pumping signal ($V_{SP}$) as a function of the bias field ($H_b$) at a constant microwave of 15 dBm, with an offset for clarity. (c) Kittel fitting of the resonance peaks to extract the saturation magnetization ($M_s$). (d) Variation in normalized $V_{SP}$ as a function of magnetic field for different microwave excitation powers, with an offset for clarity.

FIG. 3. (a – c) Switching characteristics of sample A (CoFeB) for different values of constant bias field. The microwave power is swept from 11 to 20 dBm at each bias field. (d – f) Switching characteristics of sample B (NiFe). The microwave power is swept from 5 to 30 dBm. An offset is added for clarity.

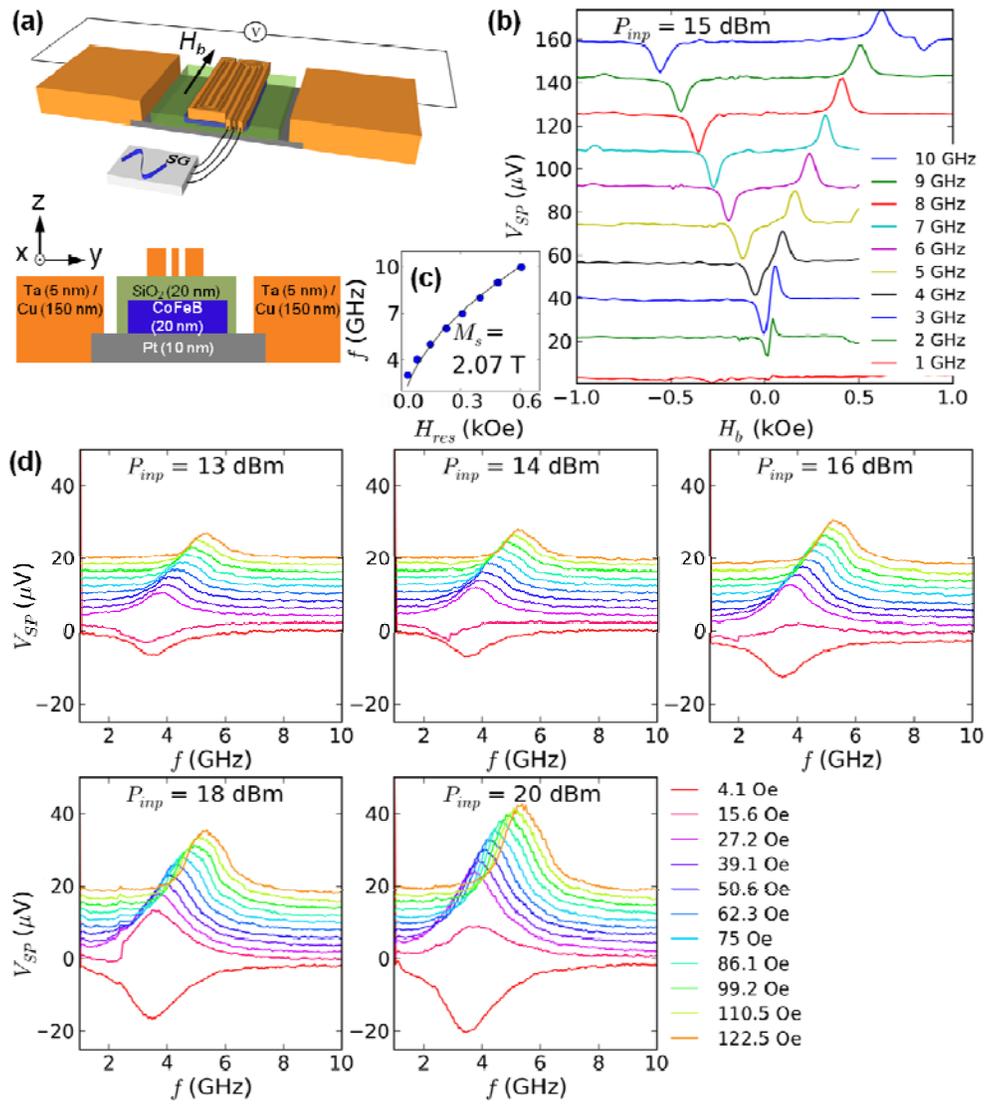

FIG. 1.

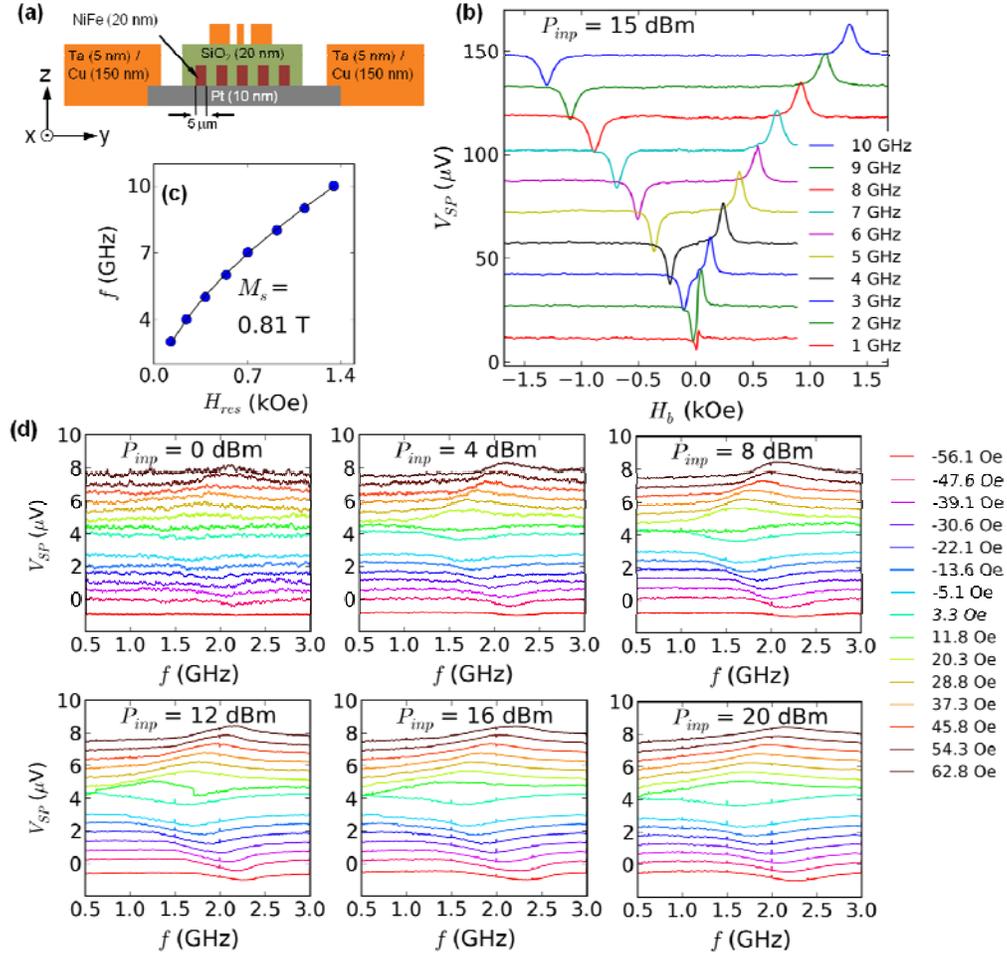

FIG. 2.

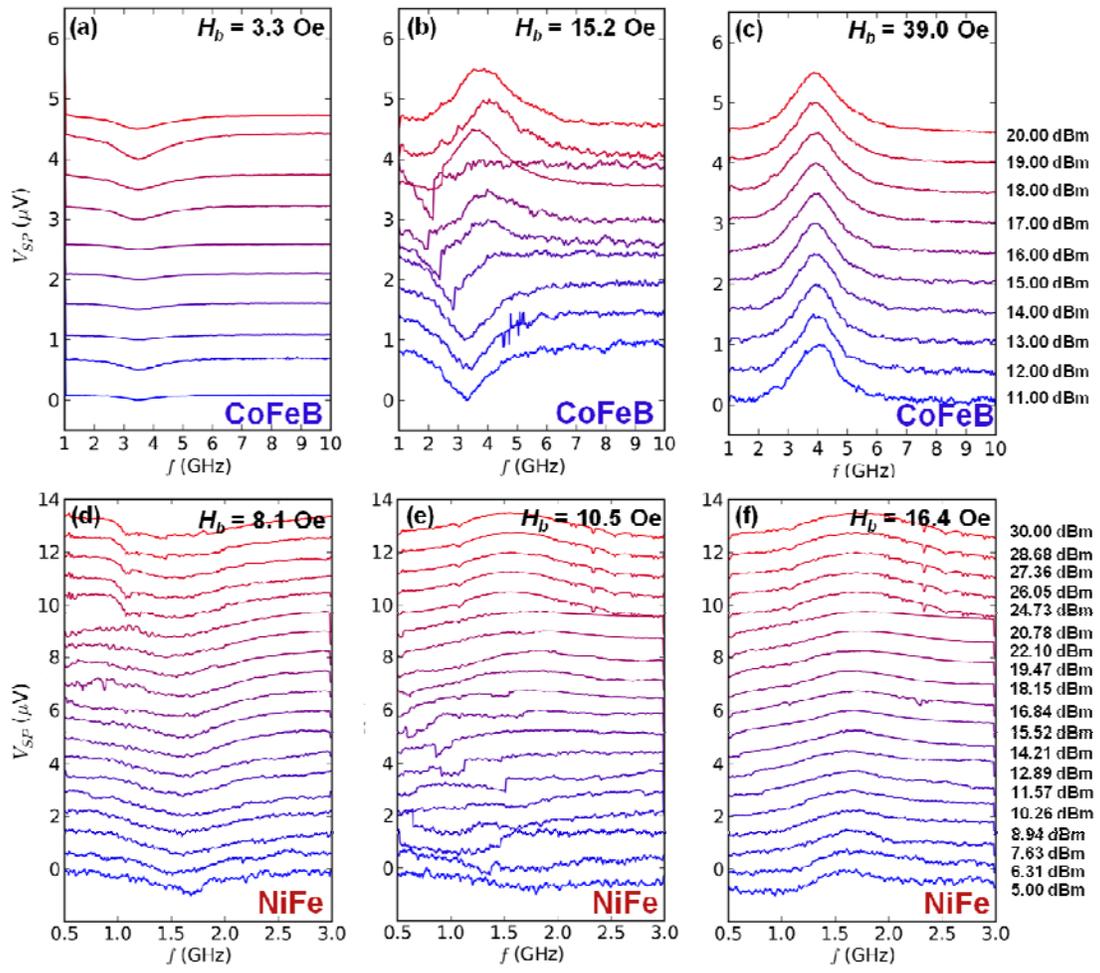

FIG. 3.